\documentclass[10pt,journal]{IEEEtran}

\usepackage{amsmath,amssymb,amsfonts}
\usepackage{algorithmic}
\usepackage{graphicx}
\usepackage{textcomp}
\usepackage{xcolor}
\usepackage{booktabs}
\usepackage{array}
\newcolumntype{L}[1]{>{\raggedright\arraybackslash}p{#1}}
\usepackage{multirow}
\usepackage{url}
\usepackage{tikz}
\usetikzlibrary{positioning,arrows.meta,calc,fit,backgrounds}
\graphicspath{{figures/}}

\definecolor{mutedblue}{RGB}{210,222,235}
\definecolor{mutedblueln}{RGB}{120,145,170}
\definecolor{mutedgreen}{RGB}{214,228,214}
\definecolor{mutedgreenln}{RGB}{120,150,120}
\definecolor{mutedamber}{RGB}{236,226,205}
\definecolor{mutedamberln}{RGB}{170,150,110}
\definecolor{mutedred}{RGB}{233,214,214}
\definecolor{mutedredln}{RGB}{170,120,120}
\definecolor{mutedgray}{RGB}{226,226,228}
\definecolor{mutedgrayln}{RGB}{140,140,145}

\tikzset{
  rabox/.style={
    rectangle, rounded corners=2pt, draw, line width=0.6pt,
    minimum width=22mm, minimum height=11mm, align=center,
    font=\footnotesize, inner sep=2pt
  },
  bxblue/.style ={rabox, fill=mutedblue,  draw=mutedblueln},
  bxgreen/.style={rabox, fill=mutedgreen, draw=mutedgreenln},
  bxamber/.style={rabox, fill=mutedamber, draw=mutedamberln},
  bxred/.style  ={rabox, fill=mutedred,   draw=mutedredln},
  bxgray/.style ={rabox, fill=mutedgray,  draw=mutedgrayln},
  raedge/.style ={-{Latex[length=2mm]}, line width=0.6pt, draw=mutedgrayln}
}

\begin{document}

\title{Risk Architecture for AI-Native Engineering Teams: An Organizational Framework for Agentic System Governance}

\author{Laxmipriya~Ganesh~Iyer\\[2pt]
{\normalfont\small Independent Researcher\\
alumni e-mail: iyer.la@northeastern.edu; ORCID: 0009-0003-7005-2527}}

\maketitle

\bstctlcite{IEEEexample:BSTcontrol}

\begin{abstract}
Engineering management research has produced mature frameworks for software risk: ownership by feature, escalation by severity, and assurance by test coverage. These frameworks implicitly assume deterministic behavior, discrete and auditable change events, and clear component-to-owner mappings. Teams that build and operate \emph{agentic} artificial intelligence (AI) systems violate all three assumptions at once: outputs are probabilistic, systems take autonomous multi-step actions, and the risk surface mutates silently between deployments. Existing AI risk literature addresses this from above (policy frameworks such as the NIST AI Risk Management Framework and ISO/IEC~42001) or from below (technical threat taxonomies such as OWASP's agentic AI guidance), but not at the layer where an engineering manager (EM) operates---the organizational layer of roles, decision rights, and escalation structures. This paper contributes (i)~a seven-dimension profile distinguishing pure software-engineering, hybrid, and AI-native teams; (ii)~a six-cluster failure-mode taxonomy that includes a previously unarticulated cluster, \emph{dependency-boundary determinism mismatch}; and (iii)~a synthetic, framework-adequacy methodology that scores how well each profile's risk architecture detects, contains, and escalates a defined scenario set. Because the object of study is framework adequacy rather than human organizational behavior, the evaluation yields derived rather than observed coverage claims. We find that coverage degrades as teams move from pure software engineering to AI-native operation---monotonically in the median, and abruptly in the count of uncovered, high-consequence failures that appear only at the AI-native step. The degradation concentrates in specific failure-mode categories, and the most severe, least-covered failures arise not inside AI-native teams but at the organizational boundary where their probabilistic outputs are consumed by determinism-assuming dependencies. We discuss implications for EM practice, change-management discipline, and the design of accountability structures for autonomous systems.
\end{abstract}


\section*{Managerial Relevance Statement}
This paper gives engineering managers a concrete, actionable way to restructure their team's risk posture for agentic AI systems---a transition for which the inherited management primitives (ownership by component, escalation by severity, assurance by test coverage) are structurally inadequate. We show, through an auditable derivation, that the highest-consequence and least-covered failures of AI-native operation are not internal to AI-native teams but arise at their organizational boundaries, where probabilistic outputs are consumed by dependencies built on deterministic assumptions---a failure mode that single-system governance and threat frameworks do not address. For practice, the analysis yields three specific shifts a manager can implement now: (i)~assign accountable ownership to \emph{surfaces}---the tool-contract layer, the causal-action chain, and the cross-team dependency boundary---not only to components; (ii)~extend escalation triggers and monitoring to \emph{semantic} signals (what an agent did, and what crossed a boundary), since conventional error- and threshold-based alerting is blind to these failures; and (iii)~design containment authority for \emph{asymmetric rollback}, including cross-boundary reconciliation when a consumer has already acted on a producer's outputs. We distill these into a minimal reference risk architecture---an ownership, trigger, and authority assignment per uncovered surface---that a manager can adopt as a starting baseline. The framework also reframes change-management discipline for a setting in which the risk surface mutates silently between deployments, replacing the change-event \emph{record} with a change-event \emph{detector}. These contributions target the engineering-management layer of roles, decision rights, and escalation structures, which sits between abstract AI-governance policy and low-level technical threat taxonomies.

\section{Introduction}
\IEEEPARstart{S}{oftware} engineering management has, over four decades, converged on a small set of durable risk-management primitives~\cite{boehm,accelerate}. Risk is owned by the team that owns the component; incidents escalate by observable severity; and assurance is established by test coverage that approximates the enumeration of system states. These primitives are not arbitrary---they are the rational response to a system model in which behavior is deterministic, change occurs through discrete and diffable events, and every consequential action is traceable to a line of code and, through it, to an owner. They also presuppose a stable alignment between system structure and team structure---a correspondence long noted in organizational theories of software, from Conway's law to modern team-topology guidance~\cite{conway,skelton}.

The emergence of \emph{agentic} AI systems---systems in which large language models (LLMs) or learned policies plan and execute multi-step action sequences with limited or no per-action human approval---disturbs each of these assumptions at once. Outputs become probabilistic: the same input need not produce the same output, and reproducibility requires frozen weights and zero-temperature sampling that do not reflect production conditions. Actions become autonomous: an agent may send communications, modify records, or invoke external services within a single run. And the risk surface becomes mutable between deployments: providers update weights, third-party tool contracts drift, and long-running agents accumulate state that alters behavior with no corresponding change event.

\begin{figure}[t]
\centering
\begin{tikzpicture}[node distance=4mm]
  \node[bxblue]  (policy) {Policy /\\ governance\\ {\scriptsize NIST, ISO, EU}};
  \node[bxred, below=of policy]  (em) {EM layer\\ {\scriptsize (this paper)}};
  \node[bxgreen, below=of em]    (tech) {Technical\\ taxonomies\\ {\scriptsize OWASP, CLTC}};
  \node[left=7mm of em, font=\scriptsize, align=right, text=mutedredln]
        (gap) {white\\ space};
  \draw[raedge] (gap.east) -- (em.west);
  \node[right=7mm of policy, font=\scriptsize, align=left] {governs the\\ \emph{system}};
  \node[right=7mm of em, font=\scriptsize, align=left, text=mutedredln]
        {governs the\\ \emph{team}};
  \node[right=7mm of tech, font=\scriptsize, align=left] {classifies\\ \emph{threats}};
\end{tikzpicture}
\caption{Three altitudes of AI risk work. Existing literature populates the policy and technical layers; the engineering-management (EM) organizational layer is the unaddressed white space.}
\label{fig:altitudes}
\end{figure}
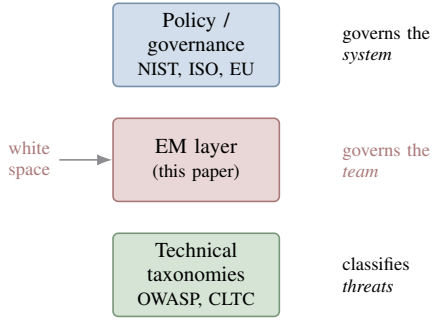

A substantial body of work has responded to AI risk, but it clusters at two altitudes. At the \emph{policy} altitude, frameworks such as the NIST AI Risk Management Framework (AI RMF)~\cite{nist}, ISO/IEC~42001~\cite{iso42001}, and the EU AI Act~\cite{euaiact} prescribe governance obligations for AI systems in the abstract. At the \emph{technical} altitude, threat taxonomies such as the OWASP agentic AI guidance~\cite{owasp} and academic risk catalogs~\cite{cltc} enumerate attack surfaces and failure classes for specific system designs. Both are valuable. Neither describes what an \emph{engineering manager} does differently---how risk ownership is assigned, how incidents are triaged and escalated, and how change is governed---when the managed team's systems are agentic. That middle layer, the organizational risk architecture, is the white space this paper addresses (Fig.~\ref{fig:altitudes}).

We argue that the EM layer is not merely an application of the policy or technical layers but a distinct unit of analysis. The relevant question is not ``how do we govern this AI system'' but ``how does an engineering manager structure their team's risk posture when the system is agentic.'' This reframing matters because several of the most consequential failure modes are organizational before they are technical: a tool's behavior drifts but no process exists to detect it, an incident occurs but no role owns it, or a probabilistic output crosses a team boundary into a dependency never designed to receive variance.

\subsection{Contributions}
This paper makes the following contributions:
\begin{enumerate}
\item \textbf{A seven-dimension team profile taxonomy} (Section~\ref{sec:profiles}) that operationalizes the distinction between \emph{pure software-engineering}, \emph{hybrid}, and \emph{AI-native} teams along output determinism, action autonomy, verification model, risk ownership, escalation trigger, data surface, and change velocity.
\item \textbf{A six-cluster failure-mode taxonomy} (Section~\ref{sec:scenarios}) covering security, privacy, autonomy, change velocity, ownership/accountability, and---newly---\emph{dependency-boundary determinism mismatch}, a cluster absent from existing single-system frameworks.
\item \textbf{A synthetic framework-adequacy methodology} (Section~\ref{sec:method}) that scores detection, containment, and escalation coverage of each team profile's risk architecture against the scenario set, including a two-team setup that models a determinism-assuming consumer of an AI-native producer's outputs.
\item \textbf{Derived findings} (Section~\ref{sec:results}) characterizing where and how coverage degrades, with the central result that the dependency boundary is the dominant source of uncovered, high-consequence failure.
\end{enumerate}

We are explicit about methodological scope. Synthetic evaluation here establishes \emph{framework adequacy}, not organizational behavior. The result is analytic rather than empirical: we do not claim that real teams behave as modeled; we claim that \emph{if} a team's risk architecture matches a given profile, \emph{then} coverage of certain failure modes is structurally determined. The value of formalizing this is that the derivation is made auditable and falsifiable---the premises are stated as code, and we show (Section~\ref{sec:robustness}) both that the findings are robust to perturbation of every modeling choice except one named premise, and that deliberately negating that premise breaks the headline finding. A derivation that could not be broken would be vacuous; ours can, and the single premise it rests on---that AI-native teams as built today lack named owners for the contract, causal-chain, and boundary surfaces---is exactly the claim the construct-validity protocol (Section~\ref{sec:construct}) and the documented-incident mapping (Section~\ref{sec:incidents}) are introduced to support.

\section{Related Work}\label{sec:related}
We organize prior work into three altitudes and show that each leaves the EM organizational layer unaddressed.

\subsection{Policy and Governance Frameworks}
The NIST AI RMF~\cite{nist} organizes AI risk management around the functions of \emph{govern}, \emph{map}, \emph{measure}, and \emph{manage}, and is deliberately system- and sector-agnostic. ISO/IEC~42001~\cite{iso42001} specifies requirements for an AI management system at the organizational level, emphasizing policy, objectives, and continual improvement. The EU AI Act~\cite{euaiact} establishes risk tiers and obligations keyed to application domain. These instruments answer \emph{what} an organization must achieve, but they are intentionally silent on the granular mechanics of how an engineering team assigns ownership, structures on-call, or triggers rollback. They treat the AI system as the unit of governance; the team is not the unit of analysis.

\subsection{Technical Threat Taxonomies}
The OWASP guidance on agentic AI~\cite{owasp}, the CLTC (UC~Berkeley) risk taxonomies~\cite{cltc}, and related academic catalogs enumerate failure and attack classes---prompt injection, tool misuse, excessive agency, privilege escalation, and others. These taxonomies are essential inputs to scenario design (we draw on them in Section~\ref{sec:scenarios}), but they classify \emph{threats to a system}, not \emph{responsibilities within a team}. They do not prescribe who owns a contract-drift incident or how it should escalate.

\subsection{Classical Software-Architecture and Organizational Risk}
A mature literature treats risk at the level of software architecture---dependency, interface, and change risk~\cite{baas,boehm}---and at the level of organizational structure, where the mapping between system decomposition and team decomposition governs how work and accountability flow~\cite{conway,skelton}. Our framework builds on both: it asks what happens to architectural risk \emph{and} its organizational ownership when components become probabilistic and agentic. The classical primitives are the baseline whose breakdown we characterize, not a target of critique.

\subsection{AI for Project and Risk Management}
A separate line of work applies AI as a \emph{tool} for risk management---predicting schedule risk, classifying defects, or summarizing incident reports. This is the inverse of our concern: we study the risk \emph{of} AI-native teams, not the use of AI to manage conventional risk.

\subsection{AI-Native Software Engineering}
Recent work on reliable and least-privilege tool-augmented agents focuses largely on system-level reliability, efficiency, and safety, and an independent line of work documents the agentic failure surface directly---e.g., indirect prompt injection that turns retrieved data into adversary-controlled instructions capable of redirecting an application's tool calls~\cite{greshake}. To our knowledge, no prior work formalizes the organizational risk architecture of AI-native teams, and in particular none articulates determinism mismatch at the dependency boundary as a first-class failure cluster. The closest technical antecedents are prior work on runtime contract verification and causal-chain gating for agentic systems~\cite{contractguard,racg}, and the agentic tool-menu risk benchmark~\cite{riskgate}, which we use as the technical substrate for the synthetic environment (Section~\ref{sec:method}). The present paper lifts those technical findings to the organizational layer: the gaps those systems close technically correspond to ownership and escalation gaps that this paper identifies organizationally.

\section{Documented-Incident Grounding}\label{sec:incidents}
Construct validity (Section~\ref{sec:construct}) establishes that the taxonomy is \emph{recognizable}; this section anchors it in the \emph{public record} by mapping each failure cluster to a documented, independently reported incident. The purpose is face validity from sources not authored for this paper: each cluster corresponds to a real, externally reported event, so the scenario set is not an internally consistent invention. We restrict this table to publicly documented incidents with first-party or authoritative third-party sources---a vendor advisory or CVE, a vendor postmortem, a regulator order, a court or tribunal ruling, or a peer-reviewable study---and we do \emph{not} rely on the author's own preprints for it (those provide the technical substrate, Section~\ref{sec:method}, not the incident evidence). Table~\ref{tab:incidents} gives the mapping.

\begin{table*}[t]
\centering
\caption{Failure Clusters Mapped to Documented Public Incidents (Independent Sources)}
\label{tab:incidents}
\renewcommand{\arraystretch}{1.3}
\small
\begin{tabular}{L{0.15\textwidth} L{0.44\textwidth} L{0.31\textwidth}}
\toprule
\textbf{Cluster} & \textbf{Documented incident (independent source)} & \textbf{Mapping to scenario} \\
\midrule
\mbox{A~Security} & \emph{EchoLeak}: zero-click prompt-injection in Microsoft 365 Copilot enabling data exfiltration from the user's context; CVE-2025-32711, Microsoft advisory (CVSS 9.3 per the Microsoft CNA; NVD base 7.5), 2025~\cite{incident-echoleak} & A1/A3: adversarial input induces an unintended agent action (data disclosure). \\
\mbox{B~Privacy} & ChatGPT March 2023 bug exposing other users' chat titles and payment details of $\sim$1.2\% of Plus subscribers; OpenAI postmortem, and the Italian Garante processing-limitation order, 2023~\cite{incident-openai-mar2023,incident-garante} & B1: sensitive data exposed at/through an external model service. \\
\mbox{C~Autonomy} & Replit AI coding agent deleted a live production database despite an explicit code freeze; reported July 2025~\cite{incident-replit} & C1/C2: irreversible action outside declared scope by a multi-step agent. \\
\mbox{D~Change} & Measured behavioral drift of GPT-3.5/GPT-4 between March and June 2023 with no transparency about updates; peer-reviewable study~\cite{incident-drift} & D1: a silent model update changes behavior with no change event. \\
\mbox{E~Ownership} & \emph{Moffatt v.\ Air Canada}: airline held liable for its chatbot's invented refund policy after arguing the bot was ``a separate legal entity''; 2024 BCCRT~149~\cite{incident-aircanada} & E1/E3: disputed/absent ownership of an agent's action. \\
\mbox{F~Boundary} & Australia's \emph{Robodebt}: an \emph{averaged} (notional) fortnightly-income estimate was consumed by an automated debt-recovery pipeline as if it were exact actual income, raising $\sim$\$1.76B in unlawful debts; Federal Court \emph{Prygodicz (No.\,2)} [2021] FCA~634 and Royal Commission Final Report, 2023~\cite{incident-robodebt-fca,incident-robodebt-rc} & F1/F3: a determinism-assuming consumer treats a probabilistic/estimated output as ground truth. \\
\bottomrule
\end{tabular}
\end{table*}

This mapping discharges the empirical burden identified in Section~\ref{sec:robustness}: the premise---that AI-native teams lack named owners for the contract, chain, and boundary surfaces---is not merely recognizable in a survey but visible in the public failure record. Two caveats keep the evidence honest. First, the boundary exemplar (Robodebt) is an \emph{arithmetic} estimate-treated-as-fact rather than a learned-model output; we use it because it is the best-evidenced public instance---located by the Federal Court precisely at the seam where an assumed figure was consumed as actual income---and because the \emph{structural} failure is identical to F1/F3 even though the generative mechanism is simpler. The same boundary failure with a learned model is exactly what Cluster~F anticipates as AI-native producers feed determinism-assuming consumers. Second, the \emph{Moffatt} case fits both C (the chatbot autonomously asserted a non-existent policy) and E (the ensuing liability dispute); we list it under E, where the accountability question is sharpest. The boundary cluster (F) remains the one least represented in existing single-system incident taxonomies and is therefore the one this paper most needs to anchor externally.

\section{Team Profile Taxonomy}\label{sec:profiles}
We characterize a team's risk-relevant operating model along seven dimensions. Each dimension takes one of three canonical values corresponding to \emph{pure SE}, \emph{hybrid}, and \emph{AI-native} profiles. A team profile is a vector of dimension values; the three canonical profiles are the extreme and intermediate points used throughout the evaluation. Table~\ref{tab:profiles} summarizes the taxonomy.

\subsection{D1: Output Determinism}
\emph{Pure SE.} Identical inputs yield identical outputs; defects are reproducible; coverage is meaningful because states can be enumerated. \emph{Hybrid.} Core business logic is deterministic while AI components (classification, ranking, summarization) are probabilistic; coverage applies only to the deterministic layer. \emph{AI-native.} Primary outputs are model-generated; reproducibility requires frozen weights and zero temperature; production behavior diverges from test behavior; coverage as a concept degrades.

\subsection{D2: Action Autonomy}
\emph{Pure SE.} The system executes explicitly coded actions; humans trigger every consequential action. \emph{Hybrid.} AI recommends; humans approve anything irreversible. \emph{AI-native.} Agents plan and execute multi-step sequences; consequential and irreversible actions can occur within a single run without per-action approval.

\subsection{D3: Test and Verification Model}
\emph{Pure SE.} Line/branch coverage, integration and regression suites, deterministic pass/fail. \emph{Hybrid.} The above plus behavioral testing of AI outputs, threshold acceptance, and human-evaluation samples. \emph{AI-native.} Behavioral coverage over input distributions, adversarial probing, red-teaming, and runtime contract verification, with probabilistic confidence bounds rather than deterministic pass/fail.

\subsection{D4: Risk and Incident Ownership}
\emph{Pure SE.} The feature team owns code, defect, and fix; on-call maps to system boundaries. \emph{Hybrid.} Ownership splits across feature, model, and data teams; incidents require cross-team triage. \emph{AI-native.} Ownership is ambiguous: a harmful autonomous action may implicate the model, the prompt author, the tool-contract owner, or the approving manager, and no standard RACI cleanly resolves it.

\subsection{D5: Escalation Trigger}
\emph{Pure SE.} Exceptions, error-rate thresholds, and SLA violations---observable, measurable, mapped to on-call. \emph{Hybrid.} The above plus model-drift detection and output-quality degradation. \emph{AI-native.} Unexpected causal action chains, runtime contract violations, in-run authorization escalation, and irreversible actions outside declared effect scope---most of which require semantic understanding of \emph{what} the agent did, not merely whether it errored, and are therefore invisible to conventional alerting.

\subsection{D6: Data Surface and Privacy Exposure}
\emph{Pure SE.} Static schema, known flows, auditable access; privacy risk bounded by schema and access control. \emph{Hybrid.} The above plus training-data lineage and inference-time exposure to external model APIs. \emph{AI-native.} A \emph{dynamic} data surface: the agent decides at runtime which tools to call and therefore which data to touch; privacy exposure depends on the causal path chosen at runtime rather than on static access configuration, and audit requires reconstructing the agent's decision path, not merely its API calls.

\subsection{D7: Change Velocity and Risk-Surface Mutation}
\emph{Pure SE.} Change is bounded, diffable, and rollback is deterministic; risk-surface changes are discrete and traceable to deployment events. \emph{Hybrid.} Code remains diffable but model updates are not; a version bump changes behavior across all inputs with no line-level diff, and rollback may not restore behavior if the model is external. \emph{AI-native.} Change is continuous and multi-layered: provider weight updates (often unannounced), low-friction prompt changes, tool-registry and third-party contract changes, and accumulated agent state. Critically, the risk surface can mutate \emph{between} deployments, violating the foundational assumption of change-management discipline that change arrives as discrete events.

\begin{table*}[t]
\centering
\caption{Seven-Dimension Team Profile Taxonomy}
\label{tab:profiles}
\renewcommand{\arraystretch}{1.3}
\small
\begin{tabular}{L{0.19\textwidth} L{0.23\textwidth} L{0.23\textwidth} L{0.23\textwidth}}
\toprule
\textbf{Dimension} & \textbf{Pure SE} & \textbf{Hybrid} & \textbf{AI-Native} \\
\midrule
D1 Output determinism & Deterministic & Mixed & Probabilistic \\
D2 Action autonomy & Human-triggered & Human-in-loop & Autonomous chains \\
D3 Test\slash verification & Coverage-based & Coverage + behavioral & Adversarial\slash contract-based \\
D4 Risk ownership & Feature team & Split\slash cross-team & Ambiguous\slash undefined \\
D5 Escalation trigger & Observable\slash measurable & + drift detection & Semantic\slash causal \\
D6 Data surface & Static\slash bounded & + model exposure & Dynamic\slash agent-determined \\
D7 Change velocity & Discrete\slash auditable & + model updates & Continuous\slash silent mutation \\
\bottomrule
\end{tabular}
\end{table*}

\subsection{Discriminant Validity of the Dimensions}\label{sec:discriminant}
A reasonable objection is whether all seven dimensions are independent, or whether some are correlates of others and a smaller set would suffice. We address the two most plausible overlaps directly. \emph{D4 (risk ownership) and D5 (escalation trigger)} are empirically correlated---ambiguous ownership often co-occurs with broken escalation---but they are conceptually distinct and can vary independently: a team may have unambiguous ownership ($O$ well-defined) yet lack any trigger capable of firing on a semantic failure ($E,M$ inadequate), and conversely may have rich triggers yet no role with authority to act. Because the rubric scores escalation against $O$ and detection against $E,M$, conflating the two would discard exactly the cases where they diverge---which are diagnostic. \emph{D6 (data surface) and D7 (change velocity)} appear to overlap for the AI-native profile because a dynamic data surface and continuous mutation often appear together, but they are orthogonal in principle: a system can have a static, bounded data surface yet high change velocity (frequent prompt and model updates over a fixed schema), or a dynamic, agent-determined surface that is nonetheless stable between rare deployments. We therefore retain seven dimensions as the minimal set that preserves these independent failure axes, and note that collapsing correlated pairs is available to practitioners who only need a coarse profile, at the cost of losing the divergent cases the rubric is designed to surface.

\section{Failure-Mode Scenario Taxonomy}\label{sec:scenarios}
Framework adequacy is measured against a defined scenario set. We organize scenarios into six clusters (A--F). Clusters A--E refine failure classes present in existing technical and governance literature into the organizational frame; Cluster~F is, to our knowledge, novel as a first-class organizational failure cluster. Each scenario is specified by (i)~a trigger, (ii)~the consequential action or exposure, and (iii)~the organizational decision point at which detection, containment, or escalation must occur.

\subsection{Cluster A: Security Failures}
\textbf{A1 Injection via AI component.} Adversarial input induces an agent to perform an unintended action. \textbf{A2 Privilege escalation.} An agent obtains or exercises authority beyond its declared scope. \textbf{A3 Unauthorized tool access.} An agent invokes a tool it should not, via misconfiguration or prompt manipulation.

\subsection{Cluster B: Privacy Failures}
\textbf{B1 PII exposure to external API.} Sensitive data (personally identifiable information, PII) is sent to an external model endpoint at inference time. \textbf{B2 Lineage loss.} The provenance of data used in a decision cannot be reconstructed. \textbf{B3 Reconstruction leakage.} Model outputs permit reconstruction of protected attributes.

\subsection{Cluster C: Autonomy Failures}
\textbf{C1 Out-of-scope irreversible action.} An agent takes an irreversible action outside its declared effect scope. \textbf{C2 Causal-chain consequence.} A sequence of individually permissible tool calls produces an unintended aggregate consequence. \textbf{C3 In-run authorization escalation.} An agent acquires elevated authority partway through a run.

\subsection{Cluster D: Change-Induced Failures}
\textbf{D1 Silent model update.} A provider weight update changes behavior with no change event. \textbf{D2 Tool-contract drift.} A tool's runtime behavior diverges from its declared contract. \textbf{D3 Prompt-change propagation.} A low-friction prompt change alters behavior across many downstream agent paths.

\subsection{Cluster E: Ownership/Accountability Failures}
\textbf{E1 Ownerless incident.} An incident occurs with no clearly accountable owner. \textbf{E2 Triage failure.} Cross-team triage cannot localize root cause. \textbf{E3 Authority gap.} An escalation trigger fires but no role holds rollback authority.

\subsection{Cluster F: Dependency-Boundary Determinism Mismatch}
This cluster captures failures that arise when an AI-native team's probabilistic outputs cross an organizational boundary into a dependency built on deterministic assumptions. The interface contract has changed---outputs are now distributional---even though the API signature has not. We model dependency consumers by their \emph{input-expectation profile}: \emph{determinism-assuming} (hard thresholds, exact-match validation, binary downstream logic, no variance tolerance), \emph{variance-aware} (confidence intervals, fallback logic, review triggers), or \emph{unaware} (built before AI-native outputs were a consideration, with variance handling simply absent).

\textbf{F1 Hard-threshold violation.} The producer emits a confidence score (e.g., $0.73$) where the consumer expects a binary label; the consumer treats the probabilistic output as certain and makes a consequential decision.

\textbf{F2 Variance cascade.} The producer's output variance is within its own acceptable bounds, but the output is piped through a chain of consumers; variance compounds until a downstream consumer receives an out-of-range input. No party in the chain owns the compounding.

\textbf{F3 Silent boundary contract drift.} The producer updates its model; the output mean is unchanged but variance increases. The API signature is unchanged, so no change notification is issued. The determinism-assuming consumer's hard-coded logic begins to fail intermittently, and root cause attribution takes weeks because the symptom presents as a downstream bug.

\textbf{F4 Rollback asymmetry.} An incident prompts the producer to roll back its model. The consumer has already processed and acted on outputs from the faulty version---records updated, communications sent, decisions logged. Producer rollback restores producer state but not consumer state; no role owns reconciliation.

\textbf{F5 Boundary ownership gap.} The consumer detects an incident and escalates to the producer. The producer's risk framework shows no anomaly (outputs were within declared bounds); the consumer's framework shows an unexpected input it does not own. The incident sits in the gap between two ownership boundaries with no resolution path.

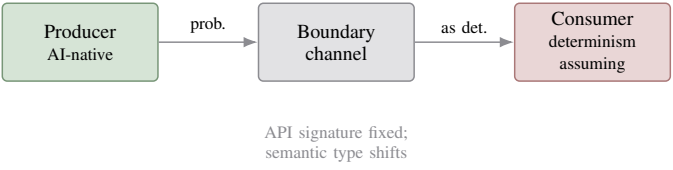
\begin{figure}[t]
\centering
\resizebox{\columnwidth}{!}{%
\begin{tikzpicture}[node distance=14mm]
  \node[bxgreen] (prod) {Producer\\ {\scriptsize AI-native}};
  \node[bxgray, right=of prod] (chan) {Boundary\\ channel};
  \node[bxred, right=of chan] (cons) {Consumer\\ {\scriptsize determinism}\\ {\scriptsize assuming}};
  \draw[raedge] (prod.east) -- node[above, font=\scriptsize] {prob.} (chan.west);
  \draw[raedge] (chan.east) -- node[above, font=\scriptsize] {as det.} (cons.west);
  \node[below=5mm of chan, font=\scriptsize, align=center, text=mutedgrayln]
        {API signature fixed;\\ semantic type shifts};
\end{tikzpicture}%
}
\caption{Cluster~F two-team setup. The static API signature on the boundary channel is unchanged, while its semantic type shifts from deterministic to probabilistic---the source of determinism-mismatch failures (F1--F5).}
\label{fig:boundary}
\end{figure}

Cluster~F is the paper's most distinctive contribution because every existing framework adopts a single-system view. The boundary view---what happens when an AI-native team's outputs are consumed by a non-AI-native dependency---is structurally outside their scope (Fig.~\ref{fig:boundary}).

\section{Methodology}\label{sec:method}
\subsection{Synthetic Environment}
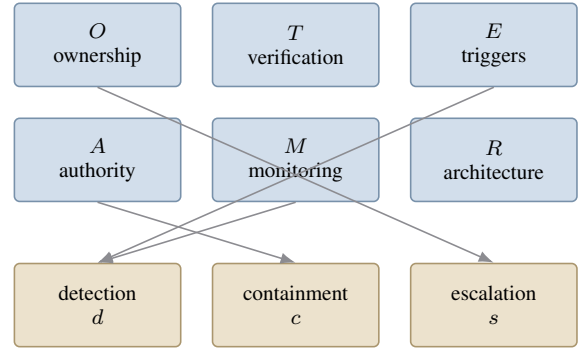
\begin{figure}[t]
\centering
\begin{tikzpicture}[node distance=4mm and 4mm]
  \node[bxblue]                 (o) {$O$\\ ownership};
  \node[bxblue, right=of o]     (t) {$T$\\ verification};
  \node[bxblue, right=of t]     (e) {$E$\\ triggers};
  \node[bxblue, below=of o]     (a) {$A$\\ authority};
  \node[bxblue, right=of a]     (m) {$M$\\ monitoring};
  \node[bxblue, right=of m]     (r) {$R$\\ architecture};
  \node[bxamber, below=8mm of a] (det) {detection\\ $d$};
  \node[bxamber, right=of det]   (con) {containment\\ $c$};
  \node[bxamber, right=of con]   (esc) {escalation\\ $s$};
  \draw[raedge] (e.south) -- (det.north);
  \draw[raedge] (m.south) -- (det.north);
  \draw[raedge] (a.south) -- (con.north);
  \draw[raedge] (o.south) -- (esc.north);
\end{tikzpicture}
\caption{Risk-architecture components ($O,T,E,A,M$) and the three capabilities they are scored on. Detection draws on triggers $E$ and monitoring $M$; containment on authority $A$; escalation on ownership $O$.}
\label{fig:tuple}
\end{figure}

We instantiate each canonical team profile as a \emph{risk architecture}: a tuple
\[
R = \langle O, \; T, \; E, \; A, \; M \rangle
\]
where $O$ is the ownership map (component or surface $\rightarrow$ accountable role), $T$ the set of verification mechanisms, $E$ the set of escalation triggers (each a predicate over observable or semantic system state), $A$ the authority map (role $\rightarrow$ permitted containment actions, e.g., rollback), and $M$ the monitoring surface (the set of events the architecture can observe). A team profile (Section~\ref{sec:profiles}) determines the contents of $R$; for example, the AI-native profile populates $E$ with semantic and causal-chain predicates and populates $M$ with a dynamic, agent-determined data surface. Fig.~\ref{fig:tuple} shows the five components and the three scored capabilities they support.

The agentic substrate---tools with risk tiers, causal action chains, authorization and escalation paths, and adversarial failure conditions---is adapted from the agentic tool-menu risk benchmark~\cite{riskgate} and its associated causal tool-filtering contracts~\cite{racg}. This provides a pre-instantiated AI-native risk surface, so the synthetic environment is grounded in an existing technical artifact rather than constructed ad hoc.

\subsection{Two-Team Boundary Setup}
To evaluate Cluster~F we instantiate a \emph{producer} (AI-native profile) and a \emph{consumer} dependency with a configurable input-expectation profile (determinism-assuming, variance-aware, or unaware). The boundary is modeled as a typed channel carrying the producer's outputs; crucially, the channel's static type (the API signature) is held fixed while its \emph{semantic} type (deterministic label vs.\ distribution) varies. Cluster~F scenarios are realized as perturbations on this channel (e.g., increased variance, model rollback) and are scored against \emph{both} architectures' $R$.

\subsection{Scoring Rubric}
For each (team profile, scenario) pair we score three capabilities on an ordinal scale $\{0,1,2\}$:
\begin{itemize}
\item \textbf{Detection} ($d$): can some $e \in E$ fire on the scenario given the monitoring surface $M$? ($0$ = no trigger exists; $1$ = a trigger exists but only post-hoc/indirect; $2$ = a direct trigger exists).
\item \textbf{Containment} ($c$): does some role in $A$ hold an authority sufficient to arrest or bound the consequence? ($0$ = none; $1$ = partial/manual; $2$ = direct).
\item \textbf{Escalation} ($s$): does ownership $O$ resolve to an accountable role with a defined path? ($0$ = ownerless/ambiguous; $1$ = resolvable via cross-team triage; $2$ = unambiguous owner).
\end{itemize}

\textbf{On the scale of measurement.} The three capability scores are \emph{ordinal}: the levels are ordered ($2 \succ 1 \succ 0$) but the spacing between them is not claimed to be uniform, so we do not treat individual scores as cardinal quantities. We therefore avoid arithmetic that would require interval assumptions on a single capability. To summarize a (profile, scenario) cell we use the \emph{coverage tier} $\tau = d + c + s \in \{0,\dots,6\}$ and report results in three bands---\textbf{Low} ($\tau \le 2$), \textbf{Medium} ($3 \le \tau \le 4$), and \textbf{High} ($\tau \ge 5$)---which are themselves ordinal and require no interval assumption. Where we aggregate across scenarios (e.g., per cluster or per profile), we report the \emph{median tier} and the count of cells in each band rather than a real-valued mean; the median is the appropriate central-tendency statistic for ordinal data. The optional normalized figure $\mathrm{cov} = \tau / 6 \in [0,1]$ is given only as a monotone visual rescaling of $\tau$ and carries no interval interpretation. Scoring is deterministic given $R$ and the scenario specification; the rubric is the framework contribution and is fully specified above so that results are reproducible and auditable rather than subjective.

\textbf{On the direct/indirect split and the band cuts.} Two thresholding choices warrant explicit justification. First, the per-capability levels $\{0,1,2\}$ distinguish \emph{no} mechanism ($0$), a mechanism that fires only \emph{post-hoc or indirectly} ($1$, e.g.\ a generic error signal that surfaces a symptom but not the cause), and a \emph{direct} mechanism aligned to the failure ($2$, e.g.\ a semantic-action or boundary-variance monitor). This three-level granularity is the coarsest that preserves the distinction the paper turns on---between a failure that is \emph{invisible} to an architecture and one it can observe only after the fact---and finer scales would invite the interval assumptions we explicitly disclaim. Second, the band cuts (Low $\tau\le2$, Medium $3$--$4$, High $\tau\ge5$) partition the tier so that \textbf{Low} denotes a cell where at most one capability is direct (or two are indirect): a configuration that cannot, by construction, both detect and own a failure. To show the headline \emph{concentration} finding is not an artifact of this cut, we re-derive it under every non-degenerate Low threshold $\tau\le k$ for $k\in\{1,2,3,4\}$ (Section~\ref{sec:robustness}): the AI-native profile has strictly more uncovered cells than either lower profile at $k\in\{2,3,4\}$, and the claim fails only at the degenerate $k=1$, where \emph{no} profile has any cell that low. The finding is therefore a property of the score distribution, not of the specific band boundary.

\subsection{Worked Example: Scoring F3}\label{sec:worked}
\begin{figure}[t]
\centering
\resizebox{\columnwidth}{!}{%
\begin{tikzpicture}[node distance=5mm and 6mm]
  \node[bxgray] (event) {F3 drift\\ {\scriptsize variance $\uparrow$}};
  \node[bxblue, right=of event] (em) {$E,M$\\ in-bounds};
  \node[bxblue, below=of em]    (a)  {$A$\\ rollback};
  \node[bxblue, below=of a]     (o)  {$O$\\ two maps};
  \node[bxred,   right=of em] (d) {$d=0$};
  \node[bxamber, right=of a]  (c) {$c=1$};
  \node[bxred,   right=of o]  (s) {$s=0$};
  \draw[raedge] (event.east) |- (em.west);
  \draw[raedge] (em.east) -- (d.west);
  \draw[raedge] (a.east)  -- (c.west);
  \draw[raedge] (o.east)  -- (s.west);
  \node[right=6mm of c, bxgray] (tau) {$\tau=1$\\ {\scriptsize Low}};
  \draw[raedge] (d.east) -- (tau.north west);
  \draw[raedge] (c.east) -- (tau.west);
  \draw[raedge] (s.east) -- (tau.south west);
\end{tikzpicture}%
}
\caption{Failure trace for F3 (determinism-assuming consumer). The drift event is invisible to in-bounds triggers $E,M$ ($d{=}0$); rollback in $A$ gives partial manual containment ($c{=}1$); ownership $O$ does not resolve across the boundary ($s{=}0$); the cell tier is $\tau=1$ (Low).}
\label{fig:trace}
\end{figure}
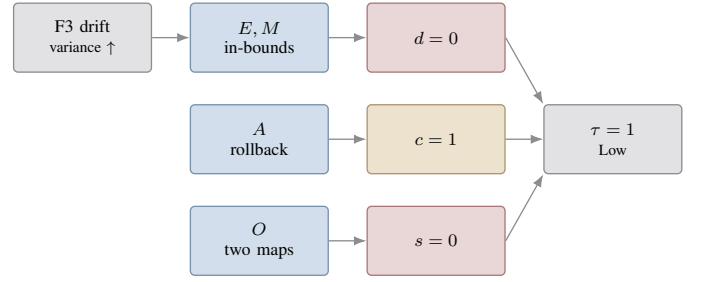

To show that the scores are \emph{derived}, not asserted, we trace scenario F3 (silent boundary contract drift) through the rubric for the producer (AI-native) architecture with a determinism-assuming consumer. In F3 the producer updates its model so that the output mean is unchanged but variance increases; the API signature is unchanged.
\begin{itemize}
\item \textbf{Detection} $d=0$. The producer's escalation triggers $E$ are predicates over its own declared output bounds, and the scenario does not violate those bounds (only second-moment behavior changed). The consumer's monitoring surface $M$ observes API-typed values, which are unchanged. No $e \in E$ on either side fires; the only signal is an intermittent downstream malfunction with no anomaly at the boundary. Hence no trigger exists.
\item \textbf{Containment} $c=1$. Once a human eventually localizes the cause, the producer holds rollback authority over its model in $A$, so the consequence can be bounded---but only manually and post-hoc, not automatically. This is partial/manual containment, not direct.
\item \textbf{Escalation} $s=0$. Ownership $O$ does not resolve: the producer's $O$ shows no anomaly (outputs were in-bounds) and the consumer's $O$ shows an unexpected input it does not own. The incident lives in the gap between the two ownership maps with no accountable role.
\end{itemize}
The resulting tier is $\tau = 0 + 1 + 0 = 1$ (band: \textbf{Low}). The same derivation applied to the variance-aware consumer raises detection to $d=1$ (confidence-band logic in the consumer's $M$ surfaces the variance shift) and escalation to $s=1$ (the consumer's framework owns the input-validation surface), giving $\tau=3$ (\textbf{Medium})---illustrating that the score difference is a property of the boundary configuration, derived directly from the contents of $R$ on each side. Fig.~\ref{fig:trace} traces this derivation against the architecture components.

\subsection{Procedure}
For each canonical profile (and, for Cluster~F, each consumer input-expectation profile), we evaluate every scenario in Clusters A--F that is \emph{in scope} for that profile---i.e.\ whose minimum complexity level the profile meets---and record $(d,c,s)$, aggregating by median tier and band as defined above. A pure-SE deterministic system cannot exhibit autonomous-chain or boundary-determinism failures, so those scenarios are marked not-applicable rather than counted as covered. The entire procedure is implemented as a small, dependency-light pipeline that encodes the profiles, the $R$ contents, the scenario requirements, and the rubric; it emits the tables and figures in Section~\ref{sec:results} directly. Because scoring is a derivation over the specified $R$ and scenarios, the procedure is fully reproducible and contains no stochastic component, and a companion test suite asserts that the reported qualitative findings hold.

\textbf{On scope and the risk of flattering the lower profiles.} The use of a per-scenario minimum complexity level ($\mathrm{min\_level}$) is consequential and warrants defense, because marking a scenario not-applicable rather than failed could artificially flatter the lower profiles---they would appear well-covered partly by being excused from the failures they would lose on. We are explicit that the scope rule \emph{is} outcome-determining here, and we defend it on semantic, not convenience, grounds. The scoping reflects a \emph{capability boundary}: a deterministic, non-agentic system cannot \emph{exhibit} an autonomous-causal-chain consequence (Cluster~C) or a probabilistic-boundary mismatch (Cluster~F) at all. Scoring a pure-SE team as ``uncovered'' on these would record a hazard to which the team is not exposed---it has no agent that can take an out-of-scope action, and no probabilistic output crossing a boundary---which measures the absence of a capability, not a risk the team carries. We verify the dependence directly rather than hide it: re-scoring with \emph{all} scenarios forced in scope for \emph{all} profiles inverts the surface picture (pure-SE accumulates the most Low cells, because it can neither detect nor own agentic failures it cannot incur), which is precisely the artifact the scope rule exists to prevent. The headline comparison is therefore explicitly conditional: \emph{among the failure modes a profile can actually exhibit}, uncovered, high-consequence failures are absent for pure-SE and hybrid and concentrate at the AI-native step. We state this conditionality plainly as a scope assumption rather than claim a scope-independent result, and we note that the \emph{within-AI-native} boundary finding (Cluster~F by consumer profile, Section~\ref{sec:results}) does not depend on the scope rule at all, since all of its cells are in scope for the AI-native producer.

\subsection{Construct Validity}\label{sec:construct}
A synthetic, derivational method raises a fair question: are the seven dimensions and the scenario set \emph{recognizable from practice}, or are they an internally consistent invention? We treat this as a question of construct validity---whether the instrument measures the intended construct---rather than of behavioral evidence, which the method explicitly does not claim. To establish construct validity we triangulate the taxonomy against three independent sources. First, the failure clusters A--E are mapped onto established external threat and governance catalogs~\cite{nist,owasp,cltc}, so that the scenario set is anchored in artifacts not authored for this paper. Second, the agentic risk surface used to instantiate the AI-native profile is taken from a pre-existing technical benchmark and its contract layer~\cite{riskgate}, rather than constructed to fit the desired result. Third, we built a structured \emph{expert-confirmation protocol} and administered it to a small panel of senior engineering managers, each asked, in structured form, (i)~whether each of the seven dimensions is recognizable in their own teams, (ii)~whether the three canonical profiles are distinguishable, and (iii)~whether each scenario corresponds to a failure they have observed or consider plausible.

We report the panel honestly and conservatively. It comprises $n=3$ senior engineering managers (median $15$ years of management experience; one director of $\sim$60 engineers) whose teams span the autonomy range the framework describes: two operate at the AI-assisted/human-in-loop tier and one ships agentic/autonomous systems. This is structured expert elicitation, not a behavioral study; it establishes that the instrument is \emph{recognizable from practice}, not how teams behave, and the sample remains small enough that we report it as supporting rather than confirmatory evidence. To keep the reader oriented to the small sample, we give each rate as a fraction of the underlying ratings rather than as a bare percentage. Across the panel, overall recognition (rating $\ge 1$) was $90/94$ ratings ($96\%$), with the profile-distinguishability and headline-finding categories recognized \emph{unanimously} ($9/9$ and $9/9$). Recognition was likewise high for the scenario set ($57/59$) and somewhat lower, but still strong, for the dimensions ($15/17$). At the recognition level, the three raters agreed on $28$ of $32$ jointly rated items; exact ordinal agreement across all three was lower ($11/33$), as expected when responses concentrate near the top of the scale. We deliberately do not report a chance-corrected agreement coefficient ($\alpha$): with recognition near-unanimous the marginals are degenerate and $\alpha$ is uninformative (a known ceiling-effect artifact), so these raw fractions are the honest summary, and we round percentages no finer than the whole number the sample supports.

Two patterns are worth surfacing rather than burying. First, the dimensions on which recognition was \emph{not} unanimous---action autonomy (D2) and dynamic data surface (D6)---were rated not-recognized only by a human-in-loop respondent, and were rated at the strongest level by the respondent shipping agentic systems. This is consistent with the framework's own thesis: recognition of the autonomy- and data-surface dimensions tracks the autonomy a team has actually shipped, rather than contradicting the taxonomy. (The earlier single-respondent pilot likewise found its two non-recognized items in the privacy cluster for a B2B e-commerce respondent, suggesting some cluster recognizability is domain-conditioned; the present panel does not yet have the spread to test that hypothesis decisively.) Second, the boundary scenarios most central to our thesis (F1/F2) were recognized at the strongest level by all three respondents, while the agreement on the headline ``high-leverage fix'' finding (Q3) came with a shared caveat---two respondents independently noted they want the variance-aware fix but ``do not see it happening''---which is itself face-valid evidence that the boundary problem is recognized as real and under-addressed. We therefore present the panel as construct-validity support for the instrument and as a demonstration that it elicits discriminating, sometimes disconfirming, responses; a larger panel with reportable chance-corrected agreement, and a separate behavioral study, remain the appropriate next steps (Section~\ref{sec:limitations}). The instrument, response schema, and scorer are released to make those a turnkey extension.

\section{Results}\label{sec:results}
We report derived coverage as ordinal bands (Section~\ref{sec:method}). All bands are consequences of the rubric applied to the specified architectures and scenarios; they are framework-adequacy results, not measurements of real teams.

\subsection{Coverage Degrades Monotonically; Uncovered Failures Appear Abruptly}
\begin{figure}[t]
\centering
\includegraphics[width=0.82\columnwidth]{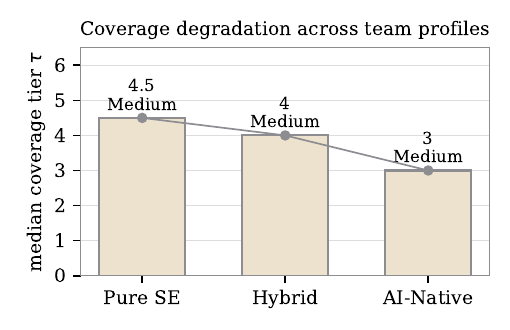}
\caption{Derived median coverage tier by team profile (computed by the released pipeline). The median declines monotonically and roughly linearly; the consequential change is the threshold appearance of uncovered (Low-band) cells at the AI-native step, where detection and escalation failures co-occur on the same scenarios.}
\label{fig:degrade}
\end{figure}

All results in this section are computed by a released, deterministic pipeline (Section~\ref{sec:method}); the figures and tables are emitted directly from that pipeline. Median coverage tiers are $\tau=4.5$ (pure SE), $\tau=4$ (hybrid), and $\tau=3$ (AI-native)---a monotone, and in the median roughly linear, decline. The median, however, is not where the consequential change appears. The telling statistic is the count of Low-band cells per profile, which is $0$ for pure SE, $0$ for hybrid, and $5$ for AI-native (Table~\ref{tab:agg}): uncovered, high-consequence failures are absent at the two lower profiles and appear abruptly at the AI-native step, where new failure surfaces (autonomy, contract drift, ownership, boundary) come into scope without a corresponding accountable owner. We are deliberate about terminology: we do not claim a non-linear decline in the median tier (it is close to linear); we claim a \emph{threshold} effect in the count of uncovered failures, concentrated at the AI-native transition. As Section~\ref{sec:robustness} shows, the dominant driver of the Low band is escalation failure ($s=0$)---an ownership gap on the new surfaces---rather than detection or containment loss; granting ownership of those surfaces alone removes every Low cell (Fig.~\ref{fig:degrade}).

\begin{table}[t]
\centering
\caption{Coverage by Team Profile (Derived, Ordinal)}
\label{tab:agg}
\renewcommand{\arraystretch}{1.3}
\begin{tabular}{l c c c c}
\toprule
\textbf{Profile} & \textbf{Median} $\tau$ & \textbf{\#Low} & \textbf{\#Med} & \textbf{\#High} \\
\midrule
Pure SE   & $4.5$ & $0$ & $2$ & $2$ \\
Hybrid    & $4$   & $0$ & $5$ & $4$ \\
AI-Native & $3$   & $5$ & $7$ & $3$ \\
\bottomrule
\end{tabular}
\end{table}

\subsection{Degradation Concentrates in Specific Clusters}
\begin{figure}[t]
\centering
\includegraphics[width=0.78\columnwidth]{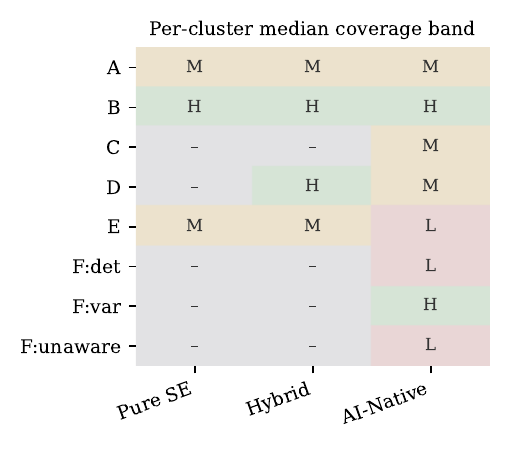}
\caption{Per-cluster median coverage band (L/M/H) by profile, emitted by the pipeline. Bands are flat for Clusters A and B but collapse across C, D, E, and the boundary cluster F; ``n/a'' denotes out-of-scope cells.}
\label{fig:heatmap}
\end{figure}

Coverage is roughly flat across profiles for Clusters A and B, but collapses for the autonomy, change, ownership, and boundary clusters as complexity rises. Several clusters are \emph{out of scope} for lower-complexity profiles---a pure-SE deterministic system cannot exhibit autonomous-chain (Cluster C) or boundary-determinism (Cluster F) failures---so those cells are marked n/a rather than scored as covered. Table~\ref{tab:cluster} and Fig.~\ref{fig:heatmap} report the per-cluster median band. The AI-native losses concentrate precisely in the categories involving autonomous action chains, causal tool sequences, runtime contract violations, and ownership (Clusters C, D2, E, F), which is the same locus addressed technically by causal-chain gating and contract-based tool filtering~\cite{racg}. The correspondence is not incidental: it is the paper's thesis that the technical mechanisms close, at the system layer, the gaps this taxonomy exposes at the organizational layer.

\begin{table}[t]
\centering
\caption{Per-Cluster Median Coverage Band (Derived, Ordinal)}
\label{tab:cluster}
\renewcommand{\arraystretch}{1.3}
\begin{tabular}{l c c c}
\toprule
\textbf{Cluster} & \textbf{Pure SE} & \textbf{Hybrid} & \textbf{AI-Native} \\
\midrule
A Security              & Medium & Medium & Medium \\
B Privacy               & High   & High   & High   \\
C Autonomy              & n/a    & n/a    & Medium \\
D Change velocity       & n/a    & High   & Medium \\
E Ownership/accountab.  & Medium & Medium & Low    \\
F Boundary (det.-assum.)& n/a    & n/a    & Low    \\
F Boundary (unaware)    & n/a    & n/a    & Low    \\
F Boundary (var.-aware) & n/a    & n/a    & High   \\
\bottomrule
\end{tabular}
\end{table}

\subsection{The Dependency Boundary Dominates Uncovered, High-Consequence Failure}
\begin{figure}[t]
\centering
\includegraphics[width=0.82\columnwidth]{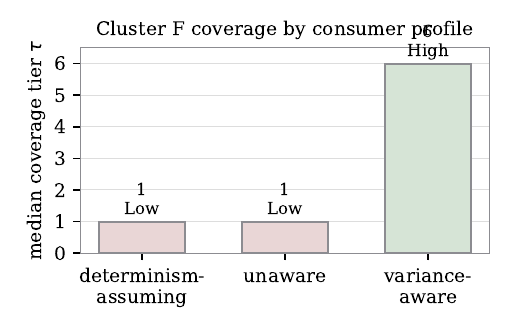}
\caption{Cluster~F median coverage tier by consumer input-expectation profile (AI-native producer held fixed). Coverage is a property of the boundary configuration: determinism-assuming and unaware consumers are Low, while a variance-aware consumer recovers High coverage.}
\label{fig:fconsumer}
\end{figure}

The lowest-covered configuration of all is Cluster~F under the determinism-assuming consumer (median band Low; F3 attains $\tau=1$). Here detection is frequently impossible ($d=0$) because the consumer's monitoring surface $M$ observes only API-typed values, which the scenario does not change (F3, F5); containment is asymmetric ($c \le 1$) because rollback authority restores only producer state, with no boundary reconciliation (F4); and escalation is ownerless ($s=0$) because no role owns the boundary channel (F5). The unaware consumer is equally Low. Under the variance-aware consumer, coverage rises to the High band---confidence-band logic in $M$ converts $d=0$ into $d\ge1$ and boundary ownership in $O$ converts $s=0$ into $s\ge1$---demonstrating that the failure is a property of the \emph{boundary configuration}, not of the producer alone (Fig.~\ref{fig:fconsumer}). This is the central practical result: the highest-consequence, least-covered failures of AI-native operation are not internal to AI-native teams but reside at their organizational boundaries with determinism-assuming dependencies.

\subsection{Robustness of the Derivation}\label{sec:robustness}
\begin{figure}[t]
\centering
\includegraphics[width=0.82\columnwidth]{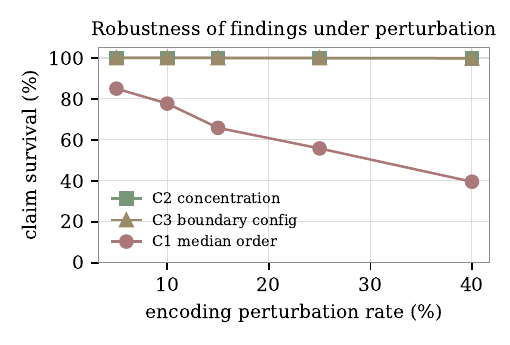}
\caption{Monte-Carlo robustness. Survival rate of each claim as the scenario-requirement encoding is randomly perturbed. The cluster-concentration and boundary-configuration findings are near-invariant; the exact pure-SE/hybrid median ordering is encoding-sensitive (a near-tie) and is reported as such.}
\label{fig:sensitivity}
\end{figure}

A derivational method invites the objection that the scores merely restate the encoding. We address this in four ways, all computed by the released pipeline. \emph{First, the result is not a tag-count artifact.} The pure-SE and AI-native architectures are encoded with the \emph{same} number of capability tags (six each; hybrid has ten), yet pure SE has zero Low-band cells and AI-native has five. Low coverage therefore cannot be explained by AI-native simply having ``fewer capabilities.'' \emph{Second, the cause is localized, not assumed.} Decomposing the AI-native Low cells, escalation failure ($s=0$, i.e.\ an ownership gap) is the dominant driver---present in four of the five---rather than an even spread across detection, containment, and escalation. As a controlled counterfactual, granting the AI-native architecture ownership of \emph{only} the three new surfaces (contract, causal chain, boundary)---adding no triggers, monitoring, or authority---removes all five Low cells. This isolates ownership of the new surfaces as the specific structural cause. \emph{Third, the concentration finding is robust to the band threshold and to perturbation of the encoding.} The choice of Low cut is not load-bearing: re-deriving the concentration claim under every non-degenerate threshold $\tau\le k$ for $k\in\{1,2,3,4\}$, AI-native has strictly more uncovered cells than either lower profile at $k\in\{2,3,4\}$ and the claim fails only at the degenerate $k=1$ (where no profile has any cell that low). We further ran a Monte-Carlo analysis that, on each of 2{,}000 trials per rate, randomly perturbs each scenario's requirements (moving tags between the direct/indirect strengths, dropping a required tag, or adding a random one) and re-derives the claims. Across perturbation rates from $5\%$ to $40\%$, the cluster-concentration and boundary-configuration claims survive in essentially all trials ($\ge 99\%$), while the exact median ordering of pure-SE versus hybrid---which depends on a near-tie ($\tau=4.5$ vs.\ $4$)---is more sensitive, surviving $86\%$ of trials at $5\%$ perturbation and degrading gracefully thereafter (Fig.~\ref{fig:sensitivity}).

\emph{Fourth, and most important, we perturb the architectures themselves and show the finding is contingent on exactly one named premise.} The objection that most threatens a derivational result is that the conclusion is encoded in the \emph{architecture} definitions---AI-native is defined without contract/chain/boundary ownership---and that perturbing only the scenario requirements (the third check above) leaves that definition untouched. We therefore added a separate harness that perturbs the capability tag-sets of the architectures and consumers directly (adding and dropping ownership, trigger/monitoring, and authority tags, type-validly, on each of 2{,}000 trials). Two conditions are compared. In the \emph{constrained} condition we hold fixed \emph{only} the Section~\ref{sec:profiles} premise---that the AI-native profile lacks named owners for the contract, causal-chain, and boundary surfaces---and perturb everything else; the boundary-configuration claim survives $\ge 95\%$ of trials and concentration $\approx 89\%$. Crucially, we then run a \emph{falsification} check: granting the AI-native architecture those three owners and nothing else \emph{breaks} the concentration finding (it no longer holds), exactly as the controlled counterfactual predicts. The headline findings are thus not unfalsifiable artifacts of calibration: they follow from a single, explicitly stated premise about how AI-native teams are organized today, they are robust to perturbation of every other modeling choice, and they are negated when that premise is negated. The empirical burden the paper carries is consequently narrow and well-localized---establish that the premise holds in practice---rather than diffuse, and it is that premise the expert protocol (Section~\ref{sec:construct}) and incident mapping (Section~\ref{sec:incidents}) address.

\subsection{Intermediate Profiles Interpolate}\label{sec:intermediate}
\begin{figure}[t]
\centering
\includegraphics[width=0.82\columnwidth]{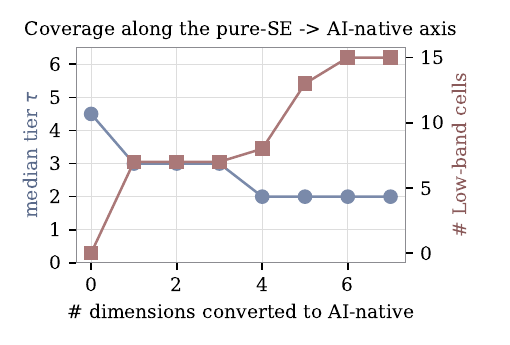}
\caption{Coverage along the pure-SE~$\rightarrow$~AI-native axis as dimensions are converted one at a time. Uncovered (Low-band) cells increase monotonically; a step appears at the first AI-native dimension, when the autonomy and boundary failure modes first come into scope.}
\label{fig:intermediate}
\end{figure}

Real teams are rarely a pure corner of the profile space: a team may be AI-native on autonomy yet pure-SE on its data surface. Because the model builds an architecture compositionally from a per-dimension level vector, we can score arbitrary mixed profiles. Converting dimensions from pure-SE to AI-native one at a time, the number of uncovered (Low-band) cells rises monotonically from $0$ (all pure-SE) to $15$ (all AI-native), and the median tier falls correspondingly (Fig.~\ref{fig:intermediate}). Two features are worth noting. The degradation is \emph{monotone}---no conversion ever \emph{improves} coverage---so intermediate teams sit between the corners as expected. But it is not smooth: there is a step as soon as the \emph{first} dimension becomes AI-native, because that is the point at which the autonomy and boundary failure modes (Clusters C and F) come into scope at all. The practical reading is that a team does not have to be fully AI-native to inherit the coverage gap; adopting agentic behavior on even one dimension activates the failure surface, while the ownership and trigger gaps accumulate as further dimensions convert.

\section{Discussion}\label{sec:discussion}
\subsection{Implications for EM Practice}
The results imply three concrete shifts in engineering-management practice for AI-native teams. First, \emph{ownership must be assigned to surfaces, not only components}: the contract layer, the causal-chain layer, and the boundary channel each require a named accountable owner, because the taxonomy shows these are precisely the surfaces where $O$ is otherwise undefined. Second, \emph{escalation triggers must include semantic predicates}: alerting that observes only errors and thresholds is structurally blind to the AI-native and Cluster~F failures, so $E$ and $M$ must be extended to observe what the agent did and what crossed the boundary. Third, \emph{authority must be designed for asymmetric rollback}: containment that restores only producer state is insufficient when consumers have already acted, so $A$ must include reconciliation authority spanning the boundary.

\subsection{A Reference Risk Architecture for AI-Native Teams}\label{sec:reference}
The three shifts above translate into a concrete starting artifact. Table~\ref{tab:reference} specifies, for each surface that the analysis identifies as uncovered, an accountable owner, a detection trigger, and a containment authority. It is deliberately minimal---a baseline an engineering manager can adapt---and it is constructed to close exactly the gaps the rubric surfaces: every row corresponds to a surface where ownership $O$, trigger $E/M$, or authority $A$ was otherwise undefined. Applying this architecture in the model (granting the new-surface ownership, semantic triggers, and reconciliation authority it prescribes) is what moves the AI-native and boundary cells out of the Low band, as the counterfactual in Section~\ref{sec:robustness} demonstrates.

\begin{table*}[t]
\centering
\caption{Reference Risk Architecture: Minimal Ownership, Trigger, and Authority Assignments for AI-Native Teams}
\label{tab:reference}
\renewcommand{\arraystretch}{1.3}
\small
\begin{tabular}{L{0.18\textwidth} L{0.20\textwidth} L{0.26\textwidth} L{0.24\textwidth}}
\toprule
\textbf{Surface} & \textbf{Accountable owner ($O$)} & \textbf{Detection trigger ($E$, $M$)} & \textbf{Containment authority ($A$)} \\
\midrule
Tool-contract layer & Contract owner (per tool/registry) & Runtime contract-violation monitor (declared vs.\ observed effect) & Disable/pin tool version; block on violation \\
Causal action chain & Agent/workflow owner & Semantic trace of executed action sequence; out-of-scope-effect alert & Kill switch on the run; quarantine the workflow \\
Dependency boundary & Boundary channel owner (jointly named producer$+$consumer) & Output-variance\slash shape monitor at the boundary; confidence-band check & Cross-boundary reconciliation authority; coordinated rollback \\
Model version & Model owner & Drift and output-quality monitor & Model rollback with downstream-impact assessment \\
Change surface (silent) & Change-surface owner & Change \emph{detector} (provider update, prompt, tool, state) rather than change record & Freeze/gate on detected mutation \\
\bottomrule
\end{tabular}
\end{table*}

\subsection{Change Management Must Become Continuous}
Dimension~D7 and Cluster~D show that change-management discipline predicated on discrete events is structurally inadequate for AI-native teams, whose risk surface can mutate between deployments. The organizational response is to treat the risk surface as continuously mutable: to monitor for silent mutation (provider updates, contract drift, state accumulation) as a first-class signal, and to define a change event \emph{detector} rather than relying on a change event \emph{record}.

\subsection{Connection to Technical Mechanisms}
The paper's findings map cleanly onto prior technical work, at two levels of abstraction. The insight from contract-based tool reliability~\cite{contractguard}---that effect integrity at the tool-contract layer is a load-bearing trust assumption---is, organizationally, the D2 ownership gap: no role owns the contract layer. The insight from causal-chain gating~\cite{racg}---that causal sequences dominate individual-tool risk---is, organizationally, the Cluster~C gap: risk frameworks govern tools, not chains. And the boundary contract-integrity result (Cluster~F) is the organizational analog of effect-integrity verification, raised from the system boundary to the team boundary. Same insight, two altitudes; together they constitute a research program rather than isolated results.

\section{Limitations and Threats to Validity}\label{sec:limitations}
\textbf{Synthetic, not behavioral.} The methodology establishes framework adequacy, not organizational behavior. We do not claim real teams behave as modeled. The claim is conditional: \emph{if} a team's risk architecture matches a profile, \emph{then} coverage of specified failure modes is structurally determined. This is a derived result and, as such, is more general than an observational survey within its scope, but it does not substitute for empirical study of adoption and human factors.

\textbf{Rubric and profile specification.} The validity of the findings rests on the team-profile definitions, the scenario set, and the scoring rubric, which we specify explicitly (Sections~\ref{sec:profiles}--\ref{sec:method}) so that they are auditable. A derivational method risks encoding its own conclusions; we therefore quantify this directly (Section~\ref{sec:robustness}). The two distinctive findings---cluster concentration and boundary-configuration dependence---are near-invariant under randomized perturbation of the encoding and are isolated to a specific structural cause (missing ownership of the contract, chain, and boundary surfaces) by a controlled counterfactual, so they are not artifacts of fine-grained calibration or of raw capability count. The weakest claim is the exact median ordering of the pure-SE and hybrid profiles, which is a near-tie and is encoding-sensitive; we report it as such rather than as a robust result.

\textbf{Scenario coverage.} The scenario set, while drawn from established threat taxonomies, is not exhaustive. Additional clusters may exist; the framework is extensible by adding scenarios and re-deriving coverage.

\textbf{Dimension coverage.} The seven dimensions characterize a team's \emph{risk-coverage} posture (detection, containment, escalation of failures), and they deliberately exclude the \emph{operational-governance} posture that surrounds AI-native development. Early expert elicitation surfaced a recurring axis not captured by D1--D7: governance and standardization of the agentic toolchain itself---cost governance and FinOps (notably token-spend minimization), security governance, and control of tool proliferation as AI coding and agentic tools multiply across a large organization. We regard this as orthogonal to, rather than a refinement of, the risk-coverage dimensions: a team may have well-defined ownership and triggers (high D4/D5) yet no standardization or cost control over the tools its agents invoke. A natural extension is an eighth, operational-governance dimension (tool-portfolio standardization, cost, and security governance) scored separately from the coverage dimensions; we leave its formalization and its interaction with the coverage results to future work.

\textbf{External validity.} The construct-validity protocol of Section~\ref{sec:construct} has been administered to a small panel of senior engineering managers ($n=3$), sufficient to establish recognizability from practice across the AI-assisted--to--agentic autonomy range but not yet to report a chance-corrected inter-rater agreement coefficient, which is uninformative at the current near-ceiling recognition rates. A larger panel ($\sim$5--10 senior EMs spanning more domains), with reportable agreement and enough variance to test the domain-conditioning hypothesis (privacy- and autonomy-cluster recognizability tracking domain and shipped autonomy), is the immediate empirical complement and remains the next step. A separate, larger study of how teams actually \emph{respond} to these failures (behavioral validity) is further future work; it would complement, not replace, the derived results.

\section{Conclusion}\label{sec:conclusion}
Engineering-management risk frameworks were designed for deterministic systems, discrete change, and component-mapped ownership. AI-native teams violate all three assumptions at once. By formalizing a seven-dimension team profile and a six-cluster failure taxonomy, and by scoring framework adequacy in a synthetic environment grounded in an existing agentic risk substrate, we show that risk coverage degrades monotonically from pure-SE to AI-native operation---with uncovered, high-consequence failures appearing abruptly at the AI-native transition---that degradation concentrates in autonomy and change-velocity failures, and that the most severe and least-covered failures arise at the organizational boundary where probabilistic outputs meet determinism-assuming dependencies. The practical consequence for engineering managers is a redesign of ownership, escalation, and authority around surfaces---contract, causal chain, and boundary---rather than around components. We position these organizational gaps as the counterparts of technical mechanisms for contract-based tool reliability and causal-chain gating, framing a unified research program spanning the system and team layers of agentic AI risk.

\section*{Data and Code Availability}
The evaluation is a deterministic derivation with no stochastic component. The pipeline that encodes the team profiles, the resource set $R$, the scenario requirements, and the scoring rubric---and that emits the tables and figures of Section~\ref{sec:results} directly---is released together with the perturbation/falsification harnesses (Section~\ref{sec:robustness}), the expert-elicitation instrument, the anonymized response schema, and the scorer (Section~\ref{sec:construct}), along with a companion test suite that asserts the reported qualitative findings hold. No human-subjects data beyond anonymized ordinal recognition ratings are included; no confidential or personally identifying information is released. The pipeline, harnesses, instrument, and test suite are available at \url{https://github.com/lakprigan/risk-architecture-ai-native-teams}.

\section*{Acknowledgment}
The author thanks the senior engineering managers who completed the construct-validity questionnaire, and colleagues whose practice informed the dimension and scenario design.

\bibliographystyle{IEEEtran}
\bibliography{references}

\end{document}